\documentstyle[preprint,aps,eqsecnum]{revtex}
\begin{document}
\title {Piezoelectric electron-phonon interaction in impure
semiconductors: 2D electrons versus composite fermions}
\author {D. V. Khveshchenko$^1$ and Michael Reizer$^2$} 

\address {$^1$ NORDITA, Blegdamsvej 17, Copenhagen DK-2100, Denmark\\
$^2$ Department of Physics, Ohio State University, Columbus, OH
43210-1106}

\maketitle

\begin{abstract}
\noindent
              
We develop a unified treatment of the piesoelectric coupling between two-dimensional electrons and bulk  
phonons in both cases of zero and strong magnetic fields, the latter 
corresponding to even denominator filling fractions.
In contrast to the case of coupling via the deformation potential,  
the leading contributions due to impurity-renormalized
electron-phonon vertices are not exactly cancelled by processes of inelastic electron-impurity scattering. Electron energy relaxation time,   
diffusion correction to the conductivity, and phonon emission rate by hot electrons 
are computed for realistic $GaAs/AlGaAs$ heterostructures.

\end{abstract}
\pagebreak

It is well known \cite{AA} that in impure metals the Coulomb interaction is
drastically enhanced as a result of a singular renormalization
of interaction vertices in the diffusive regime of small momentum and
energy transfers ( $q\ell<1$ and $\omega\tau<1$, where
$\tau$ is the electron momentum relaxation time due to elastic
electron-impurity scattering and $\ell=v_F\tau$ is the electron mean
free path). 
The earlier analysis of the effects of the deformation potential in impure metals \cite{A} 
was based on the assumption
that the electron-phonon interaction undergoes the same renormalization. 
It turned out, however, 
that the situation is essentially more complicated. 
It was shown in \cite{RS} that 
 screening of Coulomb ionic and impurity potentials
together with inelastic electron-impurity scattering leads to a
precise cancellation of the impurity renormalization of the electron-phonon
vertex. The underlying physics of this cancellation is a local
electroneutrality of the system. 

In this paper we re-examine the problem of piezoelectric electron-phonon
interactions in crystalline semiconductors 
without an inversion center, such as $GaAs$. 
In piezoelectric crystals an elastic strain is accompanied by a macroscopic
electric field, and the electron-phonon interaction results from coupling 
between electrons and the lattice polarization caused by the strain. However, 
the electroneutrality of the electron-ion
system does not affect piezoelectric coupling, so one could expect 
its effects to be different from those of the deformation
potential, particularly in the long-wavelength limit.

In a metal inelastic electron-impurity scattering is determined
by the same impurity potential $V_{e-imp}$   
which stems from a local charge disturbance. In the case of a piezoelectric crystal, however,
there is another potential source of inelastic scattering processes. 
Namely, if the size of impurity atoms is different from that of host
ones, then the lattice strain caused by this difference gives rise
to a local lattice polarization. The latter
results in an additional inelastic electron-impurity
scattering which may become important in the long-wavelength limit.  

One can argue, however, that in $Si$-doped $GaAs/AlGaAs$ 
heterostructures the inelastic electron-impurity
scattering associated with the piezoelectric coupling can be safely ignored.
Indeed, the covalent sizes of $Al$, $Ga$, $As$, and $Si$ atoms are very 
close (1.18$\AA$, 1.26$\AA$,
1.20$\AA$, and 1.19$\AA$ respectively). Therefore an additional 
perturbation induced by the impurity size appears to be much smaller than the effect 
of its Coulomb potential. Hence, 
inelastic piezoelectric electron-impurity scattering is negligible compared to
the elastic (Coulomb) one.

Recently, the interest in the problem
of the 2DEG coupled to piezoelectric phonons was boosted by 
the observation of a distinct anomaly in propagation of surface acoustic 
waves (SAW) 
in a quantizing perpendicular magnetic field corresponding to the filling factor
$\nu=1/2$ \cite{exp}. In the second part of this paper 
 we extend our zero field analysis onto the case of $\nu=1/2$ which can be readily generalized further to include 
other primary even denominator fractions (EDFs), such as $\nu=n\pm {1\over 2q}$.  

{\it Electron energy relaxation time}

In what follows we consider the range of temperatures below few Kelvin where the only important coupling
is the piesoelectric one, and treat phonons as bulk acoustic modes coupled to a local
electronic density by virtue of the bare vertex
$$M^{(0)}_{\lambda}({\bf Q})=eh_{14}(A_{\lambda}/2\rho u_{\lambda}Q)^{1/2},\\\ A_{l}={9q^2_zq^4\over 2Q^6},\\\ A_{tr}={8q^4_zq^2+q^6\over 4Q^6}      \eqno(1)$$
where ${\bf Q}=({\bf q}, q_z)$ is the 3D phonon momentum, $\rho$ is the bulk mass
density of $GaAs$, $u_{\lambda}$
is a longitudinal ($l$) or a transverse (${tr}$) sound velocity, and  $h_{14}$
is the only non-zero component of the piesoelectric tensor
(the numerical values of these and other material constants can be found, e.g., in 
the second Ref.\cite{B} which addressed the piesoelectric phonon problem in the disorder-dominated
Integer Quantum Hall regime).
  
As a result of the Coulomb interaction in the 2DEG the vertex (1) undergoes a dynamical  screening which appears to be important at practically all scales  
$$M_{\lambda}(\omega, {\bf Q})={M^{(0)}_{\lambda}({\bf Q})\over \varepsilon(\omega,q)} \eqno(2)$$
where   $\varepsilon(\omega, {q})=1+H(q){2\pi e^2\over \varepsilon_0q}P^{R}_{00}(\omega, q)$ is the 2DEG dielectric function which depends on    
the (retarded) 2D scalar polarization
 $P^{R}_{00}(\omega, q)$ and  the formfactor of the quantum well 
$H(q)=\int \int dzdz' \xi^2(z)\xi^2(z')e^{-q|z-z'|}$ given in terms of the wave function
of the lowest occupied quantum well subband $\xi(z)\sim ze^{-z/w}$.

In the diffusive regime $q\ell <1$ the dielectric function is given by the standard
formula
$$\epsilon(\omega,q)=1+{D\kappa q\over i\omega+Dq^2},              \eqno(3)$$
where $D=v_F^2\tau/2$ is the diffusion coefficient and $\kappa=2\pi e^2\nu_F/
\varepsilon_0$ is the 2D Debye wavevector proportional to the two-spin electron density of states at the Fermi level $\nu_F=m/\pi$.

As follows from the above discussion, in the diffusive regime 
one should also renormalize the vertex (2) by an
impurity ladder \cite{A}. The kinetic equation which describes electron
energy relaxation and accounts for such a renormalization was derived in \cite{A,RS}. Generalizing it onto the case of different electronic $(T)$ and 
lattice $(\Theta)$ temperatures we obtain 
$${\partial n(\epsilon)\over \partial t}=-{1\over (2\pi)^4}\sum_{\lambda}\int d^3{\bf Q}
\int d\omega |F(q_z)|^2|M_{\lambda}(\omega,{\bf Q})|^2
{\rm Im}D^R(\omega,{\bf Q})
{\rm Re}{1\over i\omega+Dq^2}R_{T,\Theta}(\epsilon,\omega)           \eqno(4)$$
where
$$R_{T,\Theta}(\epsilon,\omega)=\coth\biggl({\omega/2\Theta}\biggr)
[S(\epsilon+\omega)-S(\epsilon)]-S(\epsilon+\omega)S(\epsilon)+1,$$
$S(\epsilon)=2n(\epsilon)-1$, and $F(q_z)=\int dz e^{izq_z}\xi^2(z)$.
              
For thermal acoustic phonons with a dispersion     
 $\Omega_{\lambda}(Q)=u_{\lambda}Q$ one can use the customary expression for the
phonon propagator 
${\rm Im}D^R(\omega,Q)=-\pi [\delta(\omega-\Omega_{\lambda}(Q))-
\delta(\omega+\Omega_{\lambda}(Q))]$. 

Putting $T=\Theta$ and using the equilibrium form of the electron distribution function 
$S(\epsilon)=-\tanh(\epsilon/2T)$ we arrive at the following expression for the energy 
relaxation time 
$${1/\tau_{\epsilon}(T)}=-{\delta\over \delta n(\epsilon)}
{\partial n(\epsilon)\over \partial t}|_{\epsilon=0}=
{2\over (2\pi)^3}\sum_{\lambda}\int d^3{\bf Q}
|F(q_z)|^2|M_{\lambda}(\Omega_\lambda(Q),{\bf Q})|^2$$
$${\rm Re}{1\over i\Omega_\lambda({\bf Q})+Dq^2}
\biggl[\coth\biggl({\Omega_\lambda({\bf Q})/2T}\biggr)
-\tanh\biggl({\Omega_\lambda({\bf Q})/ 2T}\biggr)\biggr]          \eqno(5)$$
This formula is valid at temperatures below $T_1=u/\ell$
which corresponds to $q\ell\sim 1$ and characterizes the onset of the 
hydrodynamical regime (throughout this paper we neglect the difference between $u_l$ and $u_{tr}$ while making rough estimates).
 
In typically used high mobility
samples ($\mu \sim 10^{6}cm^2/Vsec$)  with sheet 
electron concentration $n_e\sim 10^{11}$cm$^{-2}$  the applicability of Eq.(5) is  
limited to the regime of extremely low temperatures ($T<T_1\sim 5mK$).

In this temperature range Eq.(5) yields 
$${1/\tau_{\epsilon}(T)}=C_3
{(eh_{14})^2\over 2\rho u_l^3}{T^2\over D\kappa^2}.              \eqno(6)$$
where the numerical factor
$$C_3={1\over 4}\int_{0}^1dx
\biggl(9x^2(1-x^2)^2+
({u_l\over u_t})^3(8x^4(1-x^2)+(1-x^2)^3)\biggr)\approx 1.35$$
receives its main contribution from two transverse phonon modes
which are slower than the longitudinal one
($u_l/u_{tr}\approx 1.73$).

Since the width of the quantum well $w\sim (\kappa k^2_F)^{-1/3}$ is typically smaller or of the order of $1/k_F$ \cite{KD}, the small-$q$ response does not probe the structure
of $\xi(z)$ and one can put the formfactors $H(q)$ and $F(q_z)$ equal to unity. 

We note, in passing, that in contrast to the conclusion drew in \cite{A} where
the case of the unscreened deformation potential was considered, the temperature dependence  of ${\tau_{\epsilon}}$ given by Eq.(5) continues all the way down to $T=0$ and 
undergoes no additional crossover at $T\sim T_2=u^2/D$ defined in \cite{A} from the condition $Dq^2\sim \Omega_{\lambda}(q)$.  
On the contrary, in our calculation of $\tau_{\epsilon}$ given by Eq.(5)
this condition does not appear at all,
since the 
factor $(i\omega +Dq^2)^{-1}$ resulting from the impurity 
vertex correction gets exactly cancelled by an identical factor
contained in $M_{\lambda}(\Omega_{\lambda}(Q), {\bf Q})$ (see Eqs.(2),(3)). 

For comparison, in the clean regime $T_1<T<T_D=2uk_F\sim 10K$ (the latter condition
facilitates that both in-plane ${\bf q}$ and out-of-plane $q_z$ components
of the phonon momentum ${\bf Q}$ are controlled by temperature
rather than by the width of the quantum well $w$) we obtain the well-known result \cite{P}:
$${1/\tau_{\epsilon}(T)}\sim 
{(eh_{14})^2\over \rho u^4}{T^3\over v_F\kappa^2}     \eqno(7)$$

Eq.(4) can be also used to determine the rate of energy transfer
from the 2DEG to the lattice  in the case of
so-called "hot" electrons with the effective temperature $T>\Theta$: 
$$P(T,\Theta)=-\int \epsilon\nu_F
{\partial n(\epsilon)\over \partial t}d\epsilon =
\sum_{\lambda}\int d^3{\bf Q}\Omega_{\lambda}(Q)
{\partial N(\Omega_{\lambda}(Q))\over \partial t}           \eqno(8)$$
The detailed derivation of the kinetic equation for the phonon distribution function  $N(\omega)$ can be found in Ref.\cite{R}.

In the strongly disordered regime we obtain the result
$$P=-{C_3\over \pi^2}{(eh_{14})^2\over 2\rho u_l^3}{\nu\over D\kappa^2}
\int^{\infty}_0 \omega d\omega
\int^{\infty}_{-\infty}\epsilon d\epsilon 
R_{T,\Theta}(\epsilon, \omega) \eqno(9)$$
The frequency integral 
$$\int^{\infty}_0 \omega d\omega
\int^{\infty}_{-\infty}\epsilon d\epsilon 
R_{T,\Theta}(\epsilon, \omega)=
2\int^{\infty}_0 [N(\omega/2\Theta)-N(\omega/2T)]\omega^3 d\omega
={2\pi^4\over 15}(\Theta^4-T^4)$$
yields a novel behavior of the electron energy loss rate
$P(T,\Theta)\sim T^4-\Theta^4$ discussed in Ref.\cite{C}, whereas in the clean limit $P(T,\Theta)$ 
follows the standard dependence ($\sim T^5-\Theta^5$).
 
As a reciprocal effect of the electron-phonon coupling, the impurity-renormalized
matrix element (2) leads to attenuation of long-wavelength phonons:  
$${1/\tau_{ph,\lambda}(u_{\lambda}Q)}= 
{|M^0_{\lambda}(\Omega_{\lambda}(Q),{\bf Q})|^2\over {2\pi e^2/\varepsilon_0q}}
{\rm Im}{1\over\varepsilon({q},\Omega_{\lambda}(Q))}=
|M_{\lambda}(\Omega_{\lambda}(Q),{\bf Q})|^2
{\rm Im}P^R_{00}({q},\Omega_{\lambda}(Q))                 \eqno(10)$$
It can be readily seen that under the conditions of the duffisive regime  $1/\tau_{ph,\lambda}(Q)$  remains constant whereas in the clean limit it 
varies as a linear function of $Q$. 

In all regimes the estimates made with the use of Eqs.(5) and (10) satisfy the relation
$${C_e(T)\over\tau_{\epsilon}(T)}\sim {C_{ph}(T)\over\tau_{ph}(T)},               \eqno(11)$$
where $C_{e,ph}(T)$ is a 2DEG ($\sim T$) or a 3D phonon ($\sim T^3$) specific heat, which is
 implied by 
the energy balance Eq.(8).

Our Eq.(9) is in agreement with the results of Ref.\cite{C} where  
the phonon emission rate was computed in the framework
of a hydrodynamical model which postulates a constancy of the conductivity 
$\sigma_{xx}(q)$ in the strongly disordered regime. 
Compared to this phenomenological approach, our microscopic analysis  
allows one to calculate  
$\tau_{\epsilon}$ and $\tau_{ph}$ directly
as well as to generalize the results onto the case of
compressible states of electrons in quantizing magnetic fields.  

{\it Conductivity correction} 

The correction to the conductivity from the piezoelectric
electron-phonon interaction can be obtained by modifying the expression derived in \cite{A}:
$${\delta\sigma_{e-ph}\over \sigma_0}=-{2\over (2\pi)^4}
\sum_{\lambda}
{\rm Im} \int d^3{\bf Q}\int d\omega D^R(\omega,{\bf Q})|M_{\lambda}(\omega,{\bf Q})|^2|F(q_z)|^2 \biggl[{Dq^2
\over (i\omega+Dq^2)^3}\biggr]f\biggl({\omega\over T}\biggr),       \eqno(12)$$
where  
$$f({\omega\over T})=
{1\over 2}\int_{-\infty}^{\infty} d\epsilon S(\epsilon+\omega)
{\partial\over \partial \epsilon}S(\epsilon)=
{\partial\over \partial \omega}
[\omega\coth{\omega\over 2T}]$$.

In the range of temperatures $T_2<T<T_1$ Eq.(12) yields
$${\delta\sigma_{e-ph}\over \sigma_0}=-{2C_1\over \pi^2}
{(eh_{14})^2\over 2\rho u_l}{1\over D^2\kappa^2}\ln\biggl({T_1\over T}\biggr),  \eqno(13)$$
where
$$C_1={1\over 4}\int_{0}^1dx\biggl(9x^2(1-x^2)+{u_l\over u_t}
(8x^4+(1-x^2)^2)\biggr]\approx 1.22$$.
At temperatures $T<T_2$ the conductivity correction $\delta\sigma_{e-ph}(T)$ ceases
to diverge logarithmically and shows only a 
quadratic deviation from its value at $T=T_2$. For the realistic parameter values
one has $T_1/T_2=u/v_F\approx 20-30$ and, hence, the interval where Eq.(13) holds is quite narrow. 

{\it Composite fermions}  

The experiments on SAW propagation \cite{exp} provide a compelling evidence that in strong magnetic fields corresponding to EDFs, such as $\nu={2\pi n_e\over B}=n\pm {1\over 2q}$
 where $n=0$ or $1$,
the ground state of the bulk 2DEG becomes gapless and    
compressible. A comprehensive description of the physics at EDFs had been achieved in the framework of the theory of composite fermions (CFs) regarded as a novel kind
of fermionic quasiparticles forming a nearly Fermi-liquid state $\cite{HLR}$.
  
The residual interactions of the CFs, as well as their interactions with charged impurities (remote
ionized donors which are usually set back some $10^2 nm$ from the 2DEG) turn out to be essentially more singular than the conventional
Coulomb potential. In the framework of the Chern-Simons theory of Ref.\cite{HLR} these interactions
are formally described as gauge forces representing local density fluctuations.

The microscopic analysis of 
the coupling between composite fermions and bulk phonons was done in \cite{KR}
where it was shown that in the presence of transverse gauge interactions 
the electron-phonon scattering is strongly enhanced compared to the zero field case
of (purely scalar) Coulomb interactions.  Experimentally, the features attributed to the enhanced electron-phonon interaction were observed in both phonon-limited electron mobility
\cite{K} and phonon drag thermopower \cite{T} at EDFs.

Namely, it was shown in \cite{KR} that 
instead of Eq.(2) the screened piezoelectric electron-phonon  matrix element
reads as (hereafter we concentrate on the case of the most prominent EDF state at $\nu=1/2$):
$$M^{cf}_{\lambda}(\omega,{\bf Q})=M^{cf,s}_{\lambda}+M^{cf,v}_{\lambda}=
{M_{\lambda}({\bf Q})\over \varepsilon_{cf}(\omega,q)}\biggl[1+ 4i\pi H(q)
{{\bf v}\times{\bf q}\over q^2}P^R_{00}(\omega,q)\biggr],            \eqno(14)$$
where $\varepsilon_{cf}=1+H(q){2\pi e^2\over 
\epsilon_0q}P^R_{00}+H(q)^2(4\pi/q)^2P^R_{00}P^R_{\perp}$.
The transverse component of the vector polarization is given by the formula 
$P^R_{\perp}(q,\omega)=\chi_{cf} q^2 +i\omega\sigma_{cf}(q)$, 
where $\chi_{cf}=1/48\pi^2\nu_{F,cf}$ and $\sigma_{cf}(q)$ equals 
${{\sqrt 2}\over 4\pi}k_Fl_{cf}$ if $ql_{cf}<2$ and ${\sqrt 2}k_{F} /(2\pi q)$ otherwise.

The diffusive regime, which is only accessible at ultra-low temperatures in the zero field
case, now sets in at much higher temperatures ($T_{1,cf}\sim 1K$), since the
bare Drude conductivity at EDFs is typically two orders of magnitude lower than 
the zero field one \cite{exp}.

One more remark is in order here.   
Recent experiments \cite{MIT} demonstrate that in the typical parameter range 
the $\nu=1/2$ state retains 
its metallic conductivity down to $T\sim 20mK$, and only goes insulating 
when the relative amount of disorder increases by a factor of two,
which can be achieved by applying a negative voltage to the side gate
and thereby depleting the 2DEG.
 In view of that we believe that in the interaction-dominated regime
the CF metal obeys a conventional diffusive behavior whereas such effects,
as a disorder-induced multifractality of single particle wave functions \cite{B}, becomes important only at excessively low temperatures. 

Proceeding along the lines of the previous zero field analysis we arrive at the formula
for the CF energy relaxation time:
$${1/\tau_{\epsilon, cf}(T)}={2\over (2\pi)^4}\sum_{\lambda}\int d^3{\bf Q}
\int d\omega{\rm Im}D^R(\omega,{\bf Q})|F(q_z)|^2
{\rm Re} \biggl[{\tau_{cf}|M^{cf,v}_{\lambda}|^2}+
{|M^{cf,s}_{\lambda}|^2\over (Dq^2+i\omega)}\biggr]$$
$$\biggl[\coth\biggl({\omega/2T}\biggr)
-\tanh\biggl({\omega/2T}\biggr)\biggr]                        \eqno(15)$$
In the disordered regime $T<T_{1,cf}$ the resulting expression for $\tau_{\epsilon, cf}$
can be cast in the form of Eq.(6) where the Debye screening momentum $\kappa_{cf}={2\pi e^2\nu_{F,cf}/\varepsilon_0}$ 
and the diffusion coefficient $D_{1/2}=\sigma_{1/2}/\nu_{F,cf}$ determined in terms of the
physical conductivity $\sigma_{1/2}\approx (e^2/2h)^2/\sigma_{cf}$ \cite{HLR} 
contain the mean-field composite fermion density of states $\nu_{F,cf}=m_{cf}/2\pi$ \cite{HLR}.

However, Eq.(6) only holds  
provided the conductivity $\sigma_{1/2}$ is large compared to 
$\sigma_M=\epsilon_0 u/2\pi\sim 10^{-2} e^2/h$. In the opposite case 
$\sigma_{1/2} < \sigma_M$ the result reads as
$${1/\tau_{\epsilon, cf}(T)}={C_5}
{(eh_{14})^2\over 2\rho u_l^5}{D_{1/2}T^2}.              \eqno(16)$$
where the value of the numerical factor
$$C_5={1\over 4}\int_{0}^1dx
\biggl(9x^2(1-x^2)^3+
({u_l\over u_t})^5(8x^4(1-x^2)^2+(1-x^2)^4)\biggr)\approx 2.48$$
is again dominated by the transverse phonon modes.

In the clean regime $T_{1,cf}<T<T_{D,cf}={\sqrt 2}T_{D}$ one obtains 
$${1/\tau_{\epsilon, cf}(T)}\sim 
{(eh_{14})^2\over \rho u^2}
{T\sigma_{1/2}k_F\over {\rm max}[u^2/D_{1/2}; D_{1/2}\kappa^2_{cf}]}                                     \eqno(17)$$
As opposed to the case of zero field described by Eqs.(6) and (7), 
the exponent in the power-law temperature dependence of $\tau_{\epsilon,cf}$ reduces by one as $T$ increases. 

Eqs.(16) and (17) are also consistent with the energy balance 
relation (11), since the phonon relaxation rate $1/\tau_{ph}(Q)$
due to scattering against CFs changes from a constant in the diffusive  
regime to the $1/Q$ behavior in the ballistic one.

In the diffusive regime both the zero field and the $\nu=1/2$ 
energy relaxation rates
share the same $T^2$ behavior. The ratio of the prefactors can be estimated as 
$${\tau^{-1}_{\epsilon,cf}\over\tau^{-1}_{\epsilon,cf}} \sim 
({\nu_F\over \nu_{F,cf}})    {\rm min} 
[ {\sigma_0\over \sigma_{1/2}};\\\ {\sigma_{0}\sigma_{1/2}\over \sigma_M^2}], \eqno(18)$$
where $\sigma_0$ is the zero field conductivity.

In pure samples exhibiting the "$\nu=1/2$ -anomaly" 
the numerical value of the right hand side of Eq.(18) is 
substantially greater than unity, 
but reduced compared to that of the transport 
relaxation rates \cite{KR} which does not contain the extra factor 
$\nu_F/\nu_{F,cf}=2m/m_{cf}\approx 0.14$ \cite{exp}. 
 
The CF conductivity correction differs from Eq.(12)
due to the fact that the CF-phonon vertex (14) contains both  scalar and 
vector parts, the latter giving
the dominant contribution:
$${\delta\sigma_{1/2}\over \sigma_{1/2}}=-{\delta\sigma_{cf}\over \sigma_{cf}}=
{\tau_{cf}\over (2\pi)^4}\sum_{\lambda}{\rm Im}\int d^3{\bf Q}
\int d\omega D^R(\omega,{\bf Q})
\biggl[{|M^{cf,v}_{\lambda}|^2\over (Dq^2+i\omega)}+
{v^2_{F,cf}q^2|M^{cf,s}_{\lambda}|^2\over (Dq^2+i\omega)^3}\biggr]
f\biggl({\omega\over T}\biggr).                                 \eqno(19)$$
The peculiar relation between the CF "quasiparticle conductivity"
$\sigma_{cf}$ and the physical current response functions \cite{HLR} 
makes both the D.C. conductivity $\sigma_{1/2}$ and the resistivity $\rho\approx \sigma_{1/2}/(e^2/2h)^2$ increase as $T\rightarrow 0$.
 
At $T_{2,cf}<T<T_{1,cf}$ and $\sigma_{1/2}>\sigma_M$ we obtain 
$${\delta\sigma_{1/2}\over \sigma_{1/2}}
={C_1\over \pi^2} {(eh_{14})^2\over \rho u_l}({2\epsilon_0\over e^2})^2
\ln\biggl({T_{1,cf}\over T}\biggr)                              \eqno(20)$$
whereas at $\sigma_{1/2}< \sigma_M$ the result reads as 
$${\delta\sigma_{1/2}\over \sigma_{1/2}}
={C_3\over \pi^2} {(eh_{14})^2\over \rho u_l^3}({\sigma_{1/2}\over \sigma_{cf}})
\ln\biggl({T_{1,cf}\over T}\biggr).                              \eqno(21)$$

The logarithmic growth of the correction
(21) stops at  $T=T_{2,cf}=u^2/D_{1/2}$ and at lower temperatures $\delta\sigma_{1/2}$ shows 
only a $\sim T^2$ deviation from its value at $T=T_{2,cf}\sim 0.1K$. 

At compressible EDF states the power $P_{cf}(T,\Theta)$ carried out 
by piesoelectric phonons from hot electrons appears to be strongly enhanced compared to that at zero field $(P_0)$.
Repeating our calculation which had led to Eq.(9)   
we obtain for $\nu=1/2$ 
$${P_{1/2}\over P_0}
=  {\rm min} 
[{\sigma_0\over \sigma_{1/2}};\\\
{C_5\over C_3}{\sigma_{0}\sigma_{1/2}\over \sigma_M^2} ] 
           \eqno(22)$$
Eq.(22) interpolates between the limiting cases of $\sigma_{1/2}$
being much greater or much smaller than $\sigma_M$. Compared to Eq.(18), the ratio (22)
is free from an
uncertainty related to a definition of the (strictly speaking, gauge-noninvariant)
CF density of states. It can be also used as an estimate
in the practically relevant case of $\sigma_{1/2}\sim \sigma_M\sim 10^{-2}e^2/h$ 
\cite{exp} where it predicts an enhancement of the rate
of phonon emission by two orders of magnitude.

It is worth mentioning that after being expressed in terms of the physical conductivity
the energy loss rate at EDFs contains absolutely no reference
to CFs used in its derivation. This formulation enables one to make a direct link to the hydrodynamical model of Ref.\cite{C} and to compare the enhancement of $P(T,\Theta)$ to that of the surface acoustic wave attenuation at $\nu=1/2$ \cite{exp}. 
  
To conclude, in the present paper we develop a unified treatment
of the piezoelectric electron-phonon interaction 
in the disordered 2DEG in both cases of zero and strong magnetic fields.
In high mobility samples 
the diffusive regime can be probed at experimentally accessible temperatures 
only in the case of strong fields corresponding to compressible EDF states. 
We compute the phonon contribution to the electron 
energy relaxation time, the power of phonon emission by hot electrons, and the conductivity correction which results from quantum interference
between electron-impurity and electron-phonon scattering.  
We observe that in the disordered regime the above quantities 
follow the same temperature dependence at zero field and at strong fields
corresponding to EDFs. In the latter case all results appear to be enhanced by the numerical factor related to the ratio of physical conductivities $\sigma_0/\sigma_{1/2}$.
 
One of the authors (M.R.) acknowledges support from US Office of Naval Research.

\end{document}